\documentclass[fleqn,usenatbib]{mnras}

\usepackage{newtxtext,newtxmath}

\usepackage[T1]{fontenc}

\DeclareRobustCommand{\VAN}[3]{#2}
\let\VANthebibliography\thebibliography
\def\thebibliography{\DeclareRobustCommand{\VAN}[3]{##3}\VANthebibliography}

\usepackage{graphicx}	
\usepackage{amsmath}	
\usepackage{multirow}

\title[The disk and the streamers of VLA 1623W]{FAUST VIII. The protostellar disk of VLA 1623--2417 W and its streamers imaged by ALMA}

\author[S. Mercimek et al.]{
S. Mercimek,$^{1,2}$\thanks{E-mail: seyma.mercimek@inaf.it}
L. Podio,$^{1}$
C. Codella,$^{1,3}$
L. Chahine,$^{4,5}$
A. L\'{o}pez-Sepulcre,$^{3,4}$
S. Ohashi,$^{6}$
L. Loinard,$^{7,8}$
\newauthor
D. Johnstone,$^{9,10}$
F. Menard,$^{3}$
N. Cuello,$^{3}$
P. Caselli,$^{11}$
J. Zamponi,$^{11}$
Y. Aikawa,$^{12}$ 
E. Bianchi,$^{13,1}$
\newauthor
G. Busquet,$^{14,15,16}$
J. E. Pineda,$^{11}$,
M. Bouvier,$^{17}$ 
M. De Simone,$^{18,1}$
Y. Zhang,$^{6,19}$
N. Sakai,$^{6}$
C. J. Chandler,$^{20}$
\newauthor
C. Ceccarelli,$^{3}$
F. Alves,$^{11}$
A. Dur\'an,$^{7}$
D. Fedele,$^{1}$
N. Murillo,$^{6}$
I. Jim\'{e}nez-Serra,$^{21}$
S. Yamamoto$^{22,23}$
\\
\\
$^{1}$INAF, Osservatorio Astrofisico di Arcetri, Largo E. Fermi 5, I-50125, Firenze, Italy\\
$^{2}$Universi\`a degli Studi di Firenze, Dipartimento di Fisica e Astronomia, via G. Sansone 1, 50019 Sesto Fiorentino, Italy\\
$^{3}$Univ. Grenoble Alpes, CNRS, IPAG, 38000 Grenoble, France\\
$^{4}$Institut de Radioastronomie Millim\'{e}trique, 38406 Saint-Martin d’H\`{e}res, France\\
$^{5}$\'Ecole doctorale de Physique, Universit\'e Grenoble Alpes, 110 Rue de la Chimie, 38400 Saint-Martin-d'H\`eres, France\\ 
$^{6}$RIKEN Cluster for Pioneering Research, 2-1, Hirosawa, Wako-shi, Saitama 351-0198, Japan\\
$^{7}$Instituto de Radioastronomía y Astrofísica , Universidad Nacional Autónoma de México, A.P. 3-72 (Xangari), 8701, Morelia, Mexico\\
$^{8}$Instituto de Astronomía, Universidad Nacional Autónoma de México, Ciudad Universitaria, A.P. 70-264, Ciudad de México 04510, Mexico\\
$^{9}$NRC Herzberg Astronomy and Astrophysics, 5071 West Saanich Road, Victoria, BC V9E 2E7, Canada\\
$^{10}$Department of Physics and Astronomy, University of Victoria, Elliott Building, 3800 Finnerty Road, Victoria, BC V8P 5C2, Canada\\
$^{11}$Max-Planck-Institut für extraterrestrische Physik (MPE), Gießenbachstr. 1, D-85741 Garching, Germany\\
$^{12}$Department of Astronomy, The University of Tokyo, 7-3-1 Hongo, Bunkyo-ku, Tokyo 113-0033, Japan\\
$^{13}$ ORIGINS, Excellence Cluster Origins, Boltzmannstrasse 2, D-85748 Garching bei München, Germany\\
$^{14}$Departament de Física Quàntica i Astrofísica, Universitat de Barcelona (UB), c/ Martí i Franquès 1, 08028 Barcelona, Spain\\
$^{15}$Institut de Ciències del Cosmos (ICCUB), Universitat de Barcelona, c. Martí i Franquès 1, 08028, Barcelona, Spain\\
$^{16}$Institut d'Estudis Espacials de Catalunya (IEEC), c. Gran Capità 2-4, 08034, Barcelona, Spain\\
$^{17}$Leiden Observatory, Leiden University, PO Box 9513, 2300 RA Leiden, The Netherlands\\
$^{18}$European Southern Observatory, Karl-Schwarzschild-Strasse 2, 85748, Garching bei München, Germany \\
$^{19}$Department of Astronomy, University of Virginia, Charlottesville, VA 22904, USA\\
$^{20}$National Radio Astronomy Observatory, PO Box O, Socorro, NM 87801, USA\\
$^{21}$Centro de Astrobiología (CSIC/INTA), Ctra.de Torrejón a Ajalvir km 4, 28806, Torrejón de Ardoz, Spain\\
$^{22}$Department of Astronomy, The University of Tokyo, 7-3-1 Hongo, Bunkyo-ku, Tokyo 113-0033, Japan\\
$^{23}$Research Center for the Early Universe, The University of Tokyo, 7-3-1 Hongo, Bunkyo-ku, Tokyo 113-0033, Japan\\
}

\date{Accepted XXX. Received YYY; in original form ZZZ}

\pubyear{2022}

\begin{document}
\label{firstpage}
\pagerange{\pageref{firstpage}--\pageref{lastpage}}
\maketitle

\begin{abstract}
More than 50\% of solar-mass stars form in multiple systems. It is therefore crucial to investigate how multiplicity affects the star and planet formation processes at the protostellar stage. We report continuum and C$^{18}$O (2--1) observations of the VLA 1623-2417 protostellar system at 50 au angular resolution as part of the ALMA Large Program FAUST. The 1.3 mm continuum probes the disks of VLA 1623A, B, and W, and the circumbinary disk of the A1+A2 binary. The C$^{18}$O emission reveals, for the first time, the gas in the disk-envelope of VLA 1623W. We estimate the dynamical mass of VLA 1623W, $M_{\rm dyn}=0.45\pm0.08$ M$_{\rm \sun}$, and the mass of its disk, $M_{\rm disk}\sim6\times10^{-3}$ M$_{\rm \sun}$. C$^{18}$O also reveals streamers that extend up to 1000 au, spatially and kinematically connecting the envelope and outflow cavities of the A1+A2+B system with the disk of VLA 1623W. The presence of the streamers, as well as the spatial ($\sim$1300 au) and velocity ($\sim$2.2 km/s) offset of VLA 1623W suggest that either sources W and A+B formed in different cores, interacting between them, or that source W has been ejected from the VLA 1623 multiple system during its formation. In the latter case, the streamers may funnel material from the envelope and cavities of VLA 1623AB onto VLA 1623W, thus concurring to set its final mass and chemical content.
\end{abstract}




\section{Introduction}




Observational studies of solar-mass star forming regions indicate a high fraction of multiplicity, $\sim30\% - 50\%$ \citep[e.g.,][]{Duchene2007, Chen2013, Tobin2016, Tobin2022, Offner2022}. 
It is, therefore, crucial to investigate what are the processes that lead
to the formation of low-mass stars and of their disks in multiple systems, in terms of dynamical interactions between protostars, ejection phenomena, and streamers stripping away or feeding gas and dust to the individual protostellar disks. 
Protostellar surveys \citep[e.g.,][]{Reipurth2000,Ward-Thompson2007,Chen2013, Tobin2016}
indicate that the multiplicity fraction (MF) is higher during the Class 0 stage (sources with age $ \sim 10^{4}$ yr, MF up to 0.5-0.8), with respect to later Class I and Class II sources (age $> 10^{5}$ yr, MF of 0.2-0.3). 
This lowering MF with time may be due to dynamical interactions that cause the ejection of one component from the protostellar system \citep{Reipurth2000,Sadavoy2017}. 
Moreover, recent interferometric observations reveal the presence of accretion streamers in protostellar systems spanning sizes from 1000 au \citep[e.g.,][]{Takakuwa2017,Alves2019, Hull2020,Pineda2022} up to $10^4$ au \citep[e.g.,][]{Pineda2020, Murillo2022}. 
Such accretion streamers are observed also at more evolved Class I/II stages and may play a crucial role in determining the final disk mass and chemical composition \citep{Garufi2022,Valvidia2022}.


The VLA 1623--2417 region (VLA 1623 hereafter) is an archetypical laboratory to investigate low-mass star formation within a multiple protostellar system and the associated phenomena: ejection, accretion, and dynamical interactions between sources \citep{Murillo2013,Harris2018,Hara2021,Ohashi2022}.
VLA1623 is located in Ophiuchus A at a distance, $d$, of 131 pc \citep{Gagne2018}, and consists of four protostellar sources: VLA 1623A, a close Class 0 binary, with separation between the components (A1 and A2) of $\sim 30$ au, surrounded by a circumbinary disk; VLA 1623B, a Class 0 protostar located $\sim130$ au West of VLA 1623A and associated with an edge-on disk; and VLA 1623W, classified as a Class I protostar and located $\sim1300$ au West of the A binary, also associated with an edge-on disk \citep[e.g.,][]{Bontemps1997,Murillo2013SED, Harris2018, Kawabe2018}. 

The A1, A2, and B protostellar sources drive high velocity outflows along the NW-SE direction detected in CO and H$_2$ \citep[e.g., ][]{Andre1990,Caratti2006,Santangelo2015,Hara2021}. The outflows open low velocity wide angle cavities observed in CCH and CS \citep[e.g.,][]{Ohashi2022}.
Emission in SO and C$^{18}$O also probes accretion flows towards the circumbinary disk of VLA 1623A \citep{Hsieh2020}. 
The outflow cavities rotate coherently with the dense parental envelope \citep{Ohashi2022}.
On the other hand, the disk of VLA 1623B counter-rotates with respect to the envelope and outflow cavities, and the A1, A2, and B disks are misaligned, suggesting
that the system is dynamically unstable \citep{Hara2021,Ohashi2022,Codella2022}. 

The nature of the more distant component of the cluster, VLA 1623W, located $\sim$1300 au away from VLA 1623A and B, and its associated phenomena are only poorly characterized. \citet{Maury2012} suggested that W could be a shocked cloudlet produced by the outflow driven by VLA 1623A. ALMA continuum images, however, revealed that W is a protostar with an edge-on disk
\citep{Harris2018,Sadavoy2019,Michel2022}. 
Based on the analysis of the spectral energy distribution (SED), \citet{Murillo2018} classified VLA 1623W as a Class I, due to the 4 times lower luminosity and 20 times lower envelope mass compared to VLA 1623A and B.
Observations of C$^{18}$O ($2-1$) line emission by \citet{Murillo2013} suggest that the systemic velocity of VLA 1623W is between 0 and 1 km\,s$^{-1}$, which differs from VLA 1623A’s systemic velocity \citep[+3.8 km\,s$^{-1}$,][]{Ohashi2022}.
\citet{Murillo2013} and \citet{Harris2018} suggest that VLA 1623W may have been ejected from the system composed by A1, A2, and B.
In this paper, we report ALMA observations of C$^{18}$O (2--1) and continuum emission at 1.3~mm used to investigate the gas in the disk of VLA 1623W, and the source dynamical interaction  with the other protostellar sources in the VLA 1623 multiple protostellar system.




\section{Observations and Data Reduction}
\label{observations}

The VLA 1623 protostellar system was observed between October 2018 and March 2020 as part of the ALMA Large Program, FAUST (Fifty AU STudy of the chemistry in the disk/envelope system of
Solar-like protostars; 2018.1.01205.L, PI: S. Yamamoto, \citealt{Codella2022}). We observed in the Band 6 frequency range 216--234\,GHz, using the 12-m array (C43-4 and C43-1) and 7-m array of the Atacama Compact Array (ACA). 
The observations were centered at $\alpha_{\rm 2000}$ = 16${^h}$26${^m}$26${^s}$.392, $\delta_{\rm 2000}$ = --24$^\circ$24$'$30$''$.69. 
The C$^{ 18}$O(2--1) line at 219560.3 MHz \citep[$E_{\rm up}$ = 16 K,][]{Muller2005}, is covered by a narrow spectral window with bandwidth of 62.5 MHz (87 km s$^{-1}$) and a channel width of 122 kHz (0.17 km s$^{-1}$).  
A spectral window with bandwidth of 1825 MHz and channel width of 977 kHz (1.25 km s$^{-1}$) has been used to image the continuum emission.

We used the Common Astronomy Software Applications package (CASA) 5.6.1-8 version \citep{McMullin2007} to obtain the calibrated visibilities, and 6.2.1 version to obtain clean, and to image the data. The multiscale deconvolver was used \citep{Cornwell2008, rau2011}.
In addition to the standard pipelines, an additional calibration routine (\url{http://www.aoc.nrao.edu/~gmoellen/}; Moellenbrock et al. in preparation) has been used to correct for the $T_\mathrm{sys}$ and for spectral data normalization. We used line-free frequencies to recover the continuum emission for each configuration and perform self-calibration. The correction of the complex gain has been derived from the self-calibration and spontaneously carried out to line visibilities of the data. Following that, to produce continuum-subtracted line data we subtracted the continuum model, derived from the self-calibration. Also the phase self-calibration technique along with long solution interval amplitudes have been used to align positions across all configurations. 
The task $tclean$ was used to obtain the image of the continuum and the datacube of the molecular emission. We adopted  Briggs weighting with a robustness parameter of $-2.0$ (uniform weighting) for the continuum to obtain the highest angular resolution (beam: 0$\farcs$42 $\times$ 0$\farcs$32, PA=$-65^\circ$). On the other hand, a robustness parameter of $0.5$ was employed for the molecular emission to optimize the signal-to-noise, consistently with the previous FAUST paper on VLA 1623 \citep{Ohashi2022}.
Finally, we applied primary beam corrections.

\begin{figure*}
	\vspace{-1cm}
\centering 
\includegraphics[width=13.0cm, angle =90]{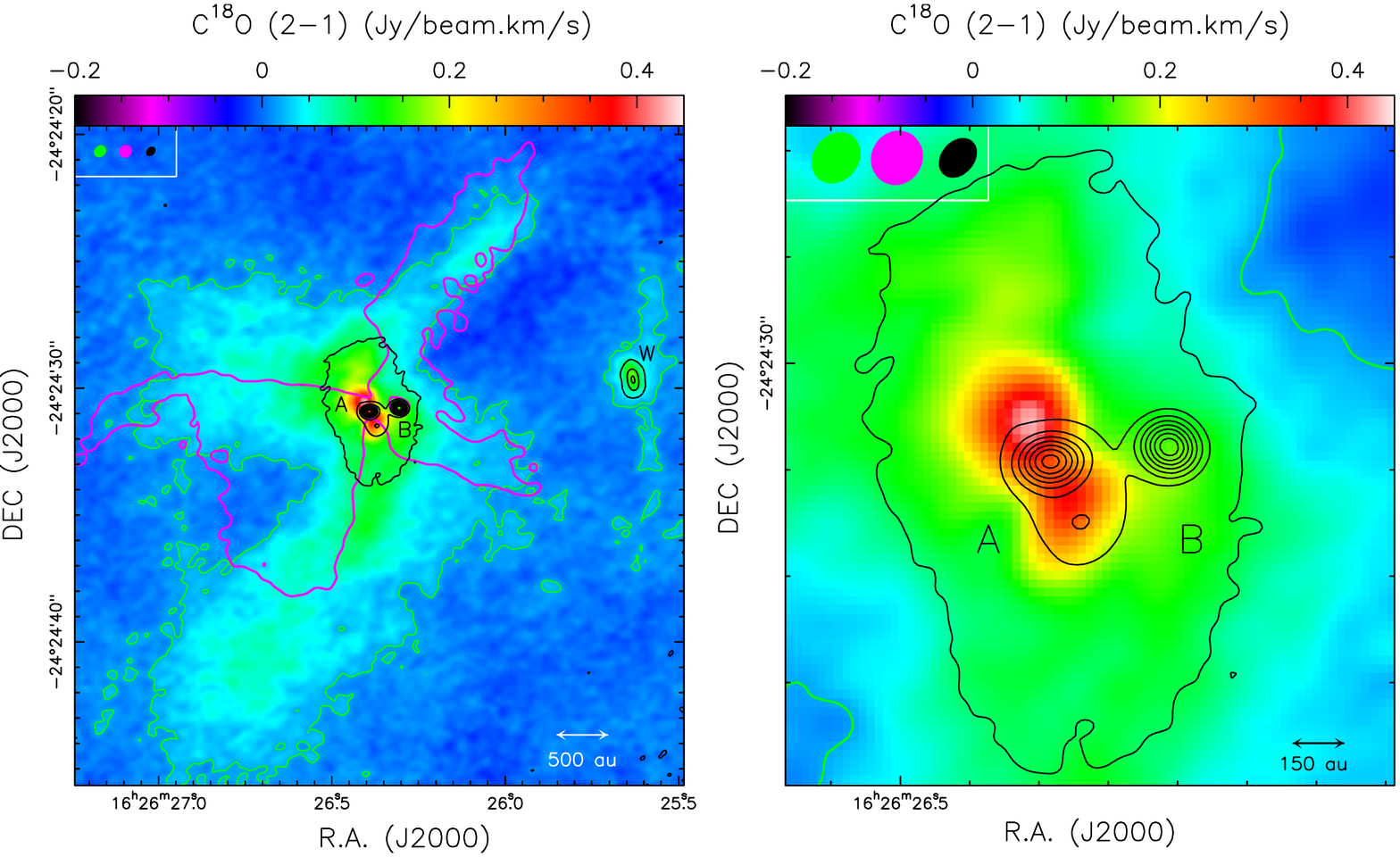}
	\vspace{-1cm}
    \caption{{\it Left:} The VLA1623--2417 system: integrated intensity map (moment 0) of C$^{18}$O (2--1) (color scale) with overlaid the dust continuum emission at 1.3 mm (black contour). The 12m+7m dataset has been used.
    The position of the A, B, and W protostars are labelled. 
    The C$^{18}$O emission is integrated from --6.0 to +10.0 km s$^{-1}$. The black continuum contours start from 3$\sigma$ (0.78 mJy beam$^{-1}$) with intervals of 40$\sigma$. The green contour is for the 3$\sigma$ level (18 mJy km s$^{-1}$ beam$^{-1}$) of the C$^{18}$O moment 0 map. 
    The magenta thick contour is the CS(5--4) emission (25$\sigma$) which traces the outflow cavity walls associated with VLA1623 A \citep[from][]{Ohashi2022}.
    The synthesized beams 
    (upper left corner) are drawn in green, magenta, and black for the C$^{18}$O, CS, and the dust continuum emission, respectively
    (Sect. \ref{observations}).
 {\it Right:} Zoom in of the moment 0 map towards sources A and B.}
    
    \label{Con-mom0}
\end{figure*}

The data analysis was carried out using the IRAM-GILDAS\footnote{\label{note5}\url{http://www.iram.fr/IRAMFR/GILDAS}} software package. 
We produced two continuum-subtracted C$^{ 18}$O (2--1) datacubes: (1) combining only the data taken with the 12-m array to sample small scales structures (beam: 0$\farcs$48 $\times$ 0$\farcs$40, PA=$-82^\circ$; $\theta_{MRS} \sim 14\arcsec$, corresponding to $\sim 1800$ au) (see Figs. \ref{channel_small} and \ref{appendix}, and panels c) and d) of Fig. \ref{supermoments}); and (2) combining the data taken with the 12-m and the 7-m arrays to recover emission extending up to $\sim 3000$ au (beam: 0$\farcs$53 $\times$ 0$\farcs$44, PA=$-74^\circ$, $\theta_{MRS} \sim 24\arcsec$, corresponding to $\sim 3150$ au) (used in Figs. \ref{Con-mom0}, \ref{channel_large}, and panels a) and b) of Fig. \ref{supermoments}). The noise root mean square (r.m.s) is 1.8 mJy beam$^{-1}$ and 1.4 mJy beam$^{-1}$ per channel for C$^{18}$O datacubes (1) and (2), respectively. The r.m.s of the continuum map is 0.26 mJy\,beam$^{-1}$.

\section{Results}
\label{results}
\subsection{Continuum emission at 1.3~mm} 
\label{Continuum}

Figure \ref{Con-mom0} shows the map of the 1.3 mm continuum emission (black contours), obtained combining the 12-m and the 7-m data. The protostellar sources A, B, and W are detected, as well as the circumbinary disk, but the angular resolution is too low to disentangle the close binary components, A1 and A2, resolved by \citet{Harris2018} at 0.9~mm (separation $\sim 30$ au). 
We fit the continuum emission towards VLA 1623A, B, and W with the CASA task \textit{imfit}, which perform a two-dimensional elliptical Gaussian fit. 
 The obtained coordinates of the continuum peak (R.A.$_{\rm J2000}$, Dec.$_{\rm J2000}$), the ellipse size, the position angle, the integrated intensity at 1.3~mm ($F_{1.3~mm}$), and the peak intensity are reported in Table \ref{Confit}. 
The results of the fit are in agreement with those obtained by \citet{Harris2018} at 0.9 mm.
For source W, we find that the disk is almost edge-on (inclination $i \simeq$ 80$^{\circ}$), and has a diameter of $\sim 93$ au. 

From the integrated intensity at 1.3~mm we estimate the mass of dust in the disks, $M_{\rm dust}$, as \citep{Hildebrand1983, Beckwith1990}:
\begin{equation}
M_{dust} = \frac{F_{1.3~mm} d^{2}}{\kappa_{\nu}B_{\nu}\left(T_{dust}\right)}.
\label{eq:quadratic}
\end{equation}
We assume isothermal conditions and optically thin emission, dust opacity ($\kappa_{\nu}$) at 1.3 mm of 2.17 cm$^{2}$g$^{-1}$ (Zamponi et al. submitted), and dust temperature, $T_{\rm dust}$, between $20$ K, the typical value assumed for Class II disks \citep[e.g.,][]{Beckwith1990}, and 50 K, to account for possible warmer dust in Class 0 and I disks \citep[e.g., ][]{Zamponi2021}. 
Table \ref{Confit} reports the derived $M_{\rm dust}$ values towards source A, source B, and source W. 

For source W, assuming a gas-to-dust ratio of 100, we obtain a total disk mass of $6\pm3 \times 10^{-3}$ M$_{\odot}$. 
The difference with the estimate obtained by \citet{Sadavoy2019} in the same ALMA band ($1 \times 10^{-2}$ M$_{\odot}$) is due to the 10\% uncertainty on the flux calibration and the different assumed distance, dust opacity and temperature.
  Moreover, the derived disk mass is affected by large uncertainty due to: (i) the assumption on the gas-to-dust ratio. If this is lower than 100 \citep[e.g., ][]{Ansdell2016}, the estimated disk mass should be regarded as an upper limit; (ii) the assumption that the dust continuum emission is optically thin. If the emission is optically thick, the estimated disk mass is a lower limit.

\begin{table*}
\centering
\begin{tabular}{lccccccc}
\hline
Source & R.A.$_{\rm J2000}$ & Dec.$_{\rm J2000}$  & Ellipse size & Position angle & $F_{\rm 1.3~mm}$ & Peak I &  M$_{\rm dust}^{*}$\\
& ($^h$ $^m$ $^s$) & ($^{\circ}$ $\arcmin$ $\arcsec$) & ($\arcsec \times \arcsec$) &  ($^{\circ}$) & (mJy) & (mJy/beam) &   ($10^{-4}$ M$_{\rm \odot}$)\\
\hline
A1+A2 &16:26:26.3907 $\pm$ 0.0013 & --24.24.30.934 $\pm$ 0.017 & 0.49 ($\pm$0.01) $\times$ 0.33 ($\pm$0.01) 
 & 74 $\pm$ 20 & 158 $\pm$ 14 & 71 $\pm$ 4   & 1.6$\pm$0.8 \\
B & 16:26:26.3063 $\pm$ 0.0001 & --24.24.30.787 $\pm$ 0.002 & 0.32 ($\pm$0.04) $\times$ 0.16 ($\pm$0.03) & 43 $\pm$ 5 & 121 $\pm$ 2 & 80 $\pm$ 1 & 1.2$\pm$0.6 \\
W &16:26:25.6312 $\pm$ 0.0002 & --24:24:29.669 $\pm$ 0.005 & 0.71 ($\pm$0.01) $\times$ 0.12 ($\pm$0.02) & 10.0 $\pm$ 0.8 & 59 $\pm$ 1 & 23.7 $\pm$ 0.3  & 0.6$\pm$0.3  \\
\hline
\end{tabular}
\caption{ Peak coordinates, ellipse size, position angle, integrated and  peak intensity of the continuum emission at 1.3~mm towards sources A1+A2, B, and W. ($^*$) The dust mass, M$_{\rm dust}$, is derived from the integrated intensity assuming a dust temperature of $20-50$ K.}
\label{Confit}
\end{table*}

\subsection{C$^{18}$O (2--1) emission}

\begin{figure*}
\centering
\includegraphics[width=12cm, angle =90]{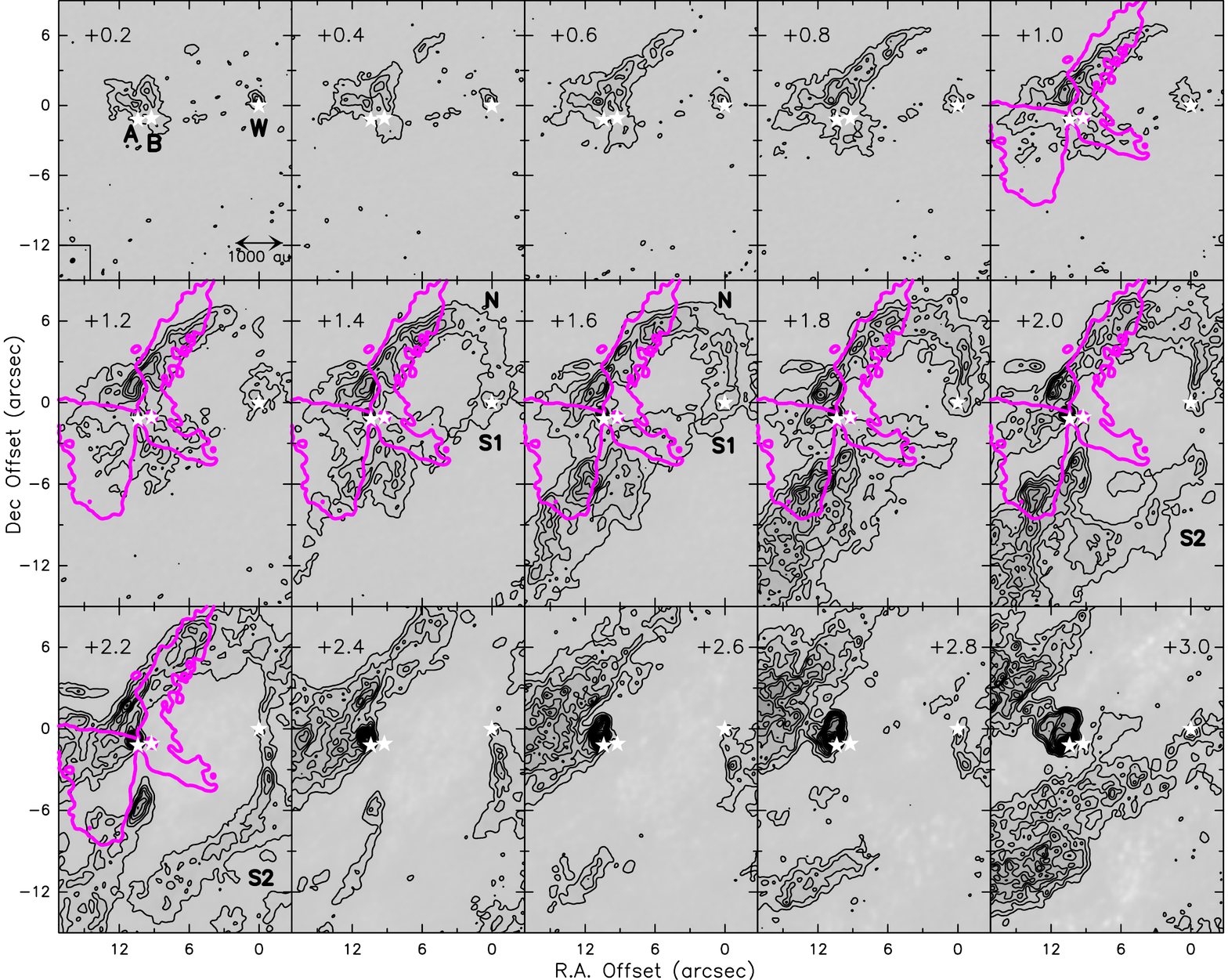}
    \caption{Channel maps of the C$^{18}$O (2--1) emission on the low velocity range ([+0.2, +3.0] km s$^{-1}$). The first contour is at 3$\sigma$ 
    (4.2 mJy beam$^{-1}$) and the step is 10$\sigma$.  The synthesized beam is shown by the black ellipse in the bottom-left corner of the first channel (beam: 0$\farcs$53 $\times$ 0$\farcs$44). The positions of VLA 1623A, B, and W are indicated by the white stars and are labelled in the first channel. The magenta contours in the channels from +1.0 km s$^{-1}$ to +2.2 km s$^{-1}$ indicate the outflow cavity walls probed by CS (5--4) emission  \citep[25$\sigma$ contour, from][]{Ohashi2022}.The northern, N, and southern, S1 and S2, streamers are labelled.}
    \label{channel_large}
\end{figure*}

\begin{figure*}
\centering
\vspace{-2cm}
	\includegraphics[width=10.5cm, angle =90]{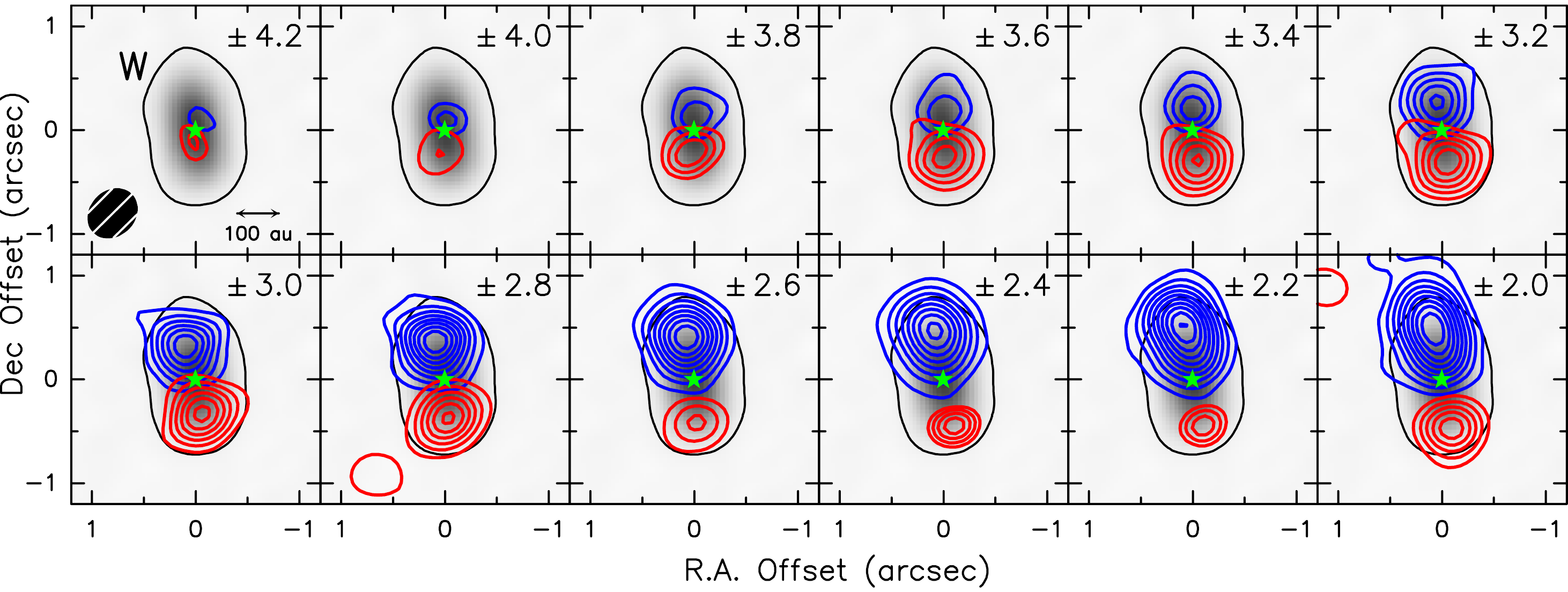}
	\vspace{-2cm}
    \caption{Channel maps of the C$^{18}$O (2--1) compact emission around the VLA 1623W protostar (green star) at high red- and blue-shifted velocities, i.e. from $\pm2$ km s$^{-1}$ to $\pm4.2$ km s$^{-1}$ with respect to the VLA 1623W systemic velocity (+1.6 km s$^{-1}$).  First contours and steps are 3$\sigma$ (5.4 mJy beam$^{-1}$). The velocity offset with respect to $V_{\rm sys}$(W) is reported in the top right corner of each panel. The black contour is the 3$\sigma$ level of the 1.3~mm continuum emission, which is also shown by the gray scale background. The synthesized beam (0$\farcs$48 $\times$ 0$\farcs$40) is shown in the bottom left corner of the first channel.}
    \label{channel_small}
\end{figure*}

Figure \ref{Con-mom0} shows the velocity-integrated intensity map (moment 0) of C$^{18}$O (2--1) towards the VLA1623 protostellar system (color scale) with overlaid the dust continuum emission at 1.3 mm (black contour) to pinpoint the positions of the protostellar sources A, B, and W, and of the circumbinary disk around the A1+A2 binary system \citep{Harris2018}. The magenta contour indicates the $25\sigma$ level of the CS(5--4) emission integrated on the velocity interval 3.4 -- 4.2 km s$^{-1}$, which probes the outflow cavity walls associated with VLA 1623A \citep[from][]{Ohashi2022}.
The C$^{18}$O emission probes the circumbinary disk and the envelope around the
A1+A2 binary system and the outflow cavity walls first identified through CS (5--4) emission \citep{Ohashi2022}. 
C$^{18}$O (2--1) also traces a bright elongated structure south of the circumbinary disk. 
Moreover, C$^{18}$O shows emission towards the VLA 1623W disk continuum, as well as an elongated structure to the North and to the South of W, roughly along the {disk position angle}. This elongated emission cannot be due to a jet or an outflow as it is not perpendicular to the disk PA.

To analyse the spatial distribution and kinematics of the different emitting components, we examine the channel maps of C$^{18}$O (see Figs. \ref{channel_large}-\ref{channel_small}). 
At low velocities, i.e. between $+0.2$  km\,s$^{-1}$ and $+3.0$ km\,s$^{-1}$,  C$^{18}$O emission extends on large scales ($>$\,$2\arcsec$ from  the continuum peak of VLA 1623W) and shows arc-like structures (hereafter called streamers) which elongate up to $>$\,$1000$ au distances from VLA 1623W connecting with the emission detected towards VLA 1623A and B (Fig. \ref{channel_large}). The spatio-kinematical properties and the origin of the C$^{18}$O  low-velocity emission is discussed in Sect. \ref{bridges}. 
In contrast, the emission at high velocities, i.e. between $-2.6$ and $-0.4$ km\,s$^{-1}$, and $+3.6$ and $+5.8$ km\,s$^{-1}$, is compact ($<1$\,$\arcsec$ from the VLA 1623W continuum peak) and shows a velocity gradient along the PA of the dusty disk (Fig. \ref{channel_small}). 
This compact high-velocity emission
probes the molecular gas in the disk of VLA 1623W and is discussed next, in Sect. \ref{inner envelope}.
\subsubsection{The gas towards the disk of VLA 1623W}
\label{inner envelope}

Figure \ref{supermoments} (panels c \& d) shows the moment 0 and moment 1 maps of C$^{18}$O (2--1) emission towards VLA 1623W integrated up to high-velocities (from --2.6 km s$^{-1}$ to +5.8 km s$^{-1}$). In this case, only the combined 12-m array is used to minimise contamination from the large scale emission. 
The continuum emission at 1.3~mm is shown by black contours. A velocity gradient along the disk PA (as derived from the continuum fit, PA$_{\rm disk} \sim 10^\circ$) is observed at a $\sim$ 50 au scale.
The moment 1 map indicates that the systemic velocity of VLA 1623W is V$_{\rm sys}$(W) $\simeq$ +1.6 km s$^{-1}$, as it corresponds to the central velocity of the range where compact blue-shifted and red-shifted emission is detected and to the mean velocity at the peak continuum emission.
In agreement with \citet{Murillo2013}, our C$^{18}$O map indicates that the systemic velocity of VLA 1623W is 
different from that of VLA 1623A and B \citep[+3.8 km s$^{-1}$,][]{Ohashi2022}.
The channel maps in Fig. \ref{channel_small} complement the information on the gas kinematics towards the disk of VLA 1623W, showing the high velocity emission  at symmetric blue-shifted and red-shifted velocities with respect to V$_{\rm sys}$(W) [from ($V_{\rm LSR} - V_{\rm sys}$) $\simeq \pm 2$ km\,s$^{-1}$ to ($V_{\rm LSR} - V_{\rm sys}$) $\simeq \pm 4.2$ km\,s$^{-1}$].  The channels maps of the emission at lower velocity, i.e. from  $V_{\rm sys}$ to $\pm 1.8$ km\,s$^{-1}$ with respect to systemic, are shown in the Appendix (Fig. \ref{appendix}), and shows that at low velocities the kinematics of the gas in the disk is contaminated by emission from the streamers mapped on larger scales in Fig. \ref{channel_large} and panels a) and b) of Fig. \ref{supermoments}. In the channel maps at high velocities, instead, the peaks of the blue- and red-shifted emission are located along the disk major axis, and the emission is more compact and peaks at smaller distances for increasing velocities with respect to V$_{\rm sys}$(W) as expected in a Keplerian rotating disk. 

The previous molecular line study towards VLA 1623W \citep{Murillo2013} showed only C$^{18}$O(2--1) blueshifted emission towards the northern disk side, plausibly due to lower sensitivity. Indeed, we find that the blue-shifted disk side is brighter than the red-shifted one up to velocities of $\pm 2.6$ km s$^{-1}$ with respect to systemic, likely due to contamination from the extended streamers observed at larger scales (see Sect. \ref{bridges}).
At higher velocities the emission from each disk side is  symmetric. We therefore use the emission in the channels at radial velocities of $\pm$ 2.8 and $\pm$ 3.0 km s$^{-1}$ to derive an estimate of the VLA 1623W dynamical mass, $M_{\rm dyn}$. The emission in these channels peak at a radial distance of 0$\farcs$37 and 0$\farcs$33. By assuming Keplerian motion we estimate a dynamical mass,
\begin{equation}
  M_{dyn} = r \frac{V^{2}}{G} 
\end{equation}
%
where $r$ and $V$ are the distance and velocity of the blue- and red-shifted peaks, deprojected for the disk inclination of 80$\degr$. The estimated dynamical mass is $0.45 \pm 0.08$ M$_{\odot}$. 

 The discovery of molecular emission towards the edge-on source W makes it a good candidate to investigate the gas vertical structure on scales $< 50$ au, as recently performed for highly inclined protoplanetary disks by \cite{Louvet2018, Teague2020, Podio2020}. 
 This is key to investigate the chemical composition of the disk in the region where planets are expected to form.
 


\subsubsection{The streamers connecting VLA 1623W with the A+B system}
\label{bridges}



\begin{figure*}
\centering
\vspace{-2cm}
	\includegraphics[width=12.5cm, angle =90]{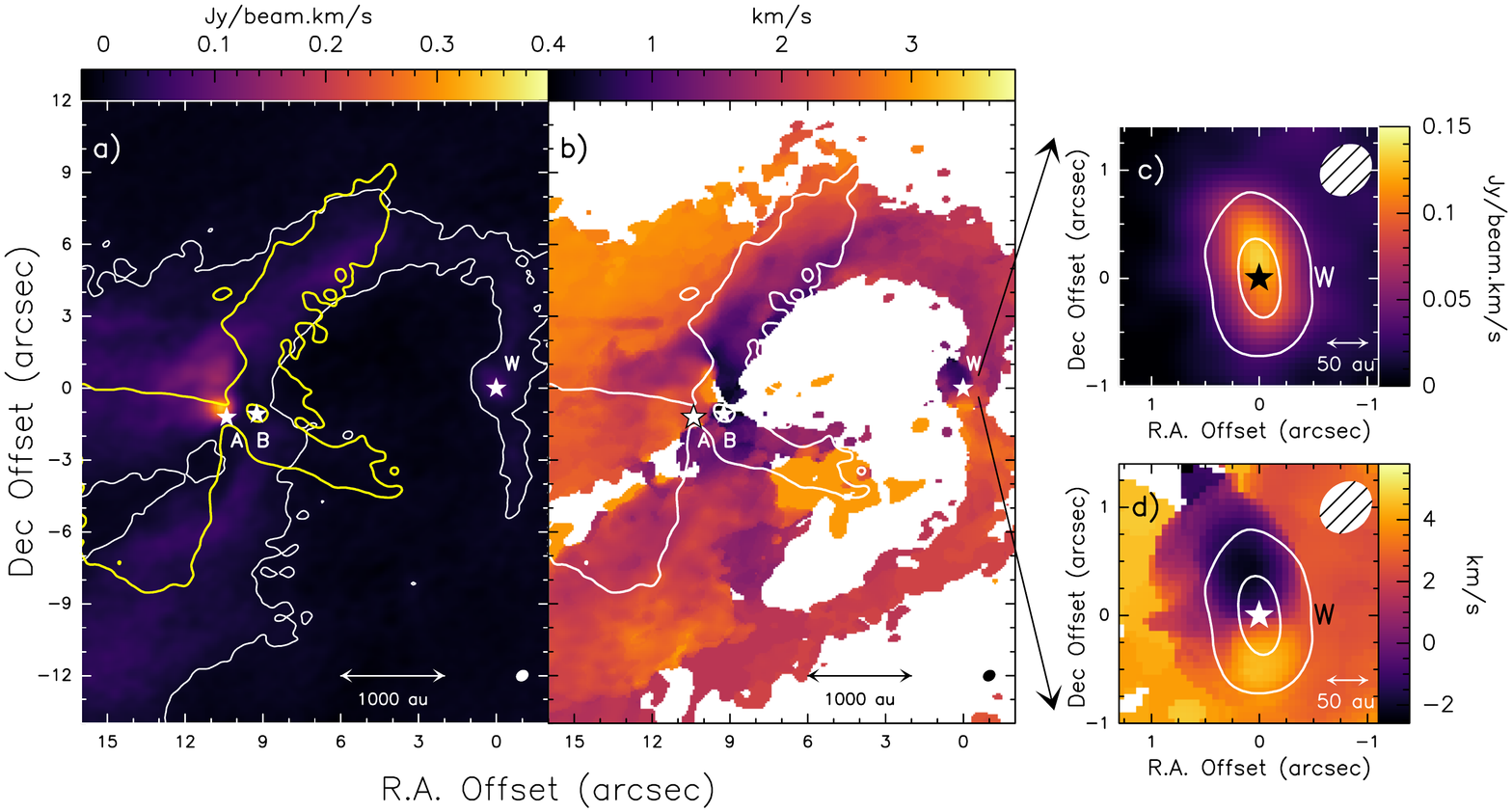}
	\vspace{-1cm}
    \caption{Integrated intensity (moment 0) and intensity-weighted mean velocity (moment 1) maps of C$^{18}$O (2--1) towards VLA 1623--2417. The RA and Dec offsets are with respect to the position of VLA 1623W. \textit{Panels a) and b)}: Moment 0 and 1 maps over the low velocity range ([+0.2, +3.0] km s$^{-1}$). 
    The white stars indicate the A, B, and W protostars, the white contour in Panel a) indicates the 3$\sigma$ emission (9 mJy km s$^{-1}$ beam$^{-1}$), the ellipse shows
    the synthesized beam (0$\farcs$53 $\times$ 0$\farcs$44), and the yellow in Panel a) and the white contour in Panel b) reveal the 25$\sigma$ CS (5--4) emission probing the cavity outflow walls associated with VLA 1623A \citep[from][]{Ohashi2022}. \textit{Panels c) and d)}: Moment 0 and moment 1 maps of VLA 1623W over the high velocity range ([--2.6, +5.8] km s$^{-1}$). Continuum emission at 1.3~mm is shown by the white contours. The black and white stars indicate the position of VLA 1623W, and the white ellipse the synthesized beam (0$\farcs$48 $\times$ 0$\farcs$40).}
    \label{supermoments}
\end{figure*}

Figure \ref{channel_large} shows the channel maps of C$^{18}$O (2--1) emission at low velocities, i.e., between +0.2 km s$^{-1}$ and +3.0 km s$^{-1}$. Three streamers connecting VLA 1623A and B with VLA 1623W are observed: one in the northern VLA1623 region, labeled as N, the other two in the southern region labeled as S1 and S2 and detected on velocities of [$+1.2$, $+1.8$] km s$^{-1}$, and [$+2.0$, $+2.8$] km s$^{-1}$, respectively.
Figure \ref{supermoments} shows the moment 0 (panel a) and moment 1 (panel b) maps obtained for the low velocity range ([+0.2, +3.0] km s$^{-1}$). Both figures use the 12-m + 7-m dataset.
The northern streamer partially overlaps with the North-West blueshifted cavity wall opened by the
outflow(s) driven by VLA 1623A \citep[see the yellow/white contours, from][]{Ohashi2022}. 
At distances from VLA 1623A larger than $\sim$ 1000 au, the molecular emission bends towards the south until it connects to the northern side of the VLA 1623W disk.
In the southern region, the streamer S1 overlaps with the South-West cavity wall opened by the outflow(s) driven by VLA 1623A. The streamer S2, on the other hand, connects the envelope surrounding VLA 1623A and B with the southern side of the VLA 1623W disk extending towards the South. In summary, the S1 and S2 streamers have both different spatial distribution and different velocities, therefore they are labelled as different streamers.

The velocities of the observed streamers are consistent with those of the outflow cavities and the envelope probed by CS(5--4) and H$^{13}$CO$^{+}$ emission by \citet[][see their Fig. 12, 13, and 16]{Ohashi2022}, and with the velocities of the VLA 1623W disk: the blueshifted outflow cavity/streamer north of VLA1623A connects with the northern blueshifted side of the VLA 1623W edge-on disk, while the redshifted outflow cavity/streamer south of VLA 1623A connects with the southern redshifted side of the VLA 1623W disk. This indicates that the protostellar sources A and W are kinematically linked. 
In addition, the moment 1 map of  C$^{18}$O shows a velocity gradient along the southern streamer having larger redshifted velocities  (by $\sim$ 1 km s$^{-1}$) at the connection with the VLA 1623W disk  with respect to the portion of streamer connected with the VLA 1623A+B envelope. 
On the contrary, no clear velocity gradient is observed along the northern streamer. The lack of a velocity gradient along the northern streamer may indicate that the gas motion occurs in the plane of the sky \citep{Alves2020}.

\section{Discussion: On the origin of VLA 1623W}
\label{Discussion}

The FAUST ALMA observations of the multiple system VLA 1623 reveal for the first time the gas kinematics towards the more distant component, VLA 1623W. Specifically, C$^{18}$O (2-1) emission probes the following structures: (i) the molecular gas in the disk of VLA 1623W with Keplerian motion on $50-100$ au scales, constraining the protostellar mass (M$_{dyn}\sim0.45$ M$_{\odot}$); (ii) streamers which extend on scales $> 1000$ au and connect spatially and kinematically the two sides of the edge-on disk of VLA 1623W (PA$_{\rm disk} \sim 10^\circ$, $i\sim 80^\circ$, $M_{\rm disk} \sim 6\times10^{-3}$ M$_{\odot}$) with the envelope associated with VLA 1623A and B.

Molecular streamers have been observed in other multiple systems, e.g., between the components of IRAS 16293-2422 separated by a distance of $\simeq$ 400 au \citep{Pineda2012,Jacobsen2018,vanderWiel2019,Murillo2022}.
In the case of VLA 1623, however, the large spatial ($\sim 1300$ au) and kinematic ($\sim 2.2$ km\,s$^{-1}$) offset between VLA 1623AB and VLA 1623W suggest that VLA 1623W does not belong to the same molecular core as VLA 1623AB. On the other hand, the molecular streamers connecting the disk of VLA 1623W with the envelope/cavities of VLA 1623A and VLA1623B suggest that VLA 1623W is not a background or foreground object.
In this context, there are two possible scenarios: (1) VLA 1623W  has formed in a core which is close by and gravitationally interacting with the envelope of VLA 1623AB; (2) VLA 1623W was ejected from the multiple system composed by A1, A2, and B due to dynamical interactions during the system’s formation, as first proposed by \citet{Murillo2013}. 

\subsection{Hypothesis 1: formation in a separate core}

In the first scenario the observed streamers are produced by the interaction of multiple cores in the Ophiuchus A star-forming region hosting the VLA 1623A, B and W protostars. 
\citet{Chen2018} mapped Ophiuchus A at spatial resolution $\geq$ 5$\arcsec$ to investigate the physical and chemical properties of the region in continuum and molecular lines by combining interferometric (SMA) and single-dish (IRAM-30m) data. The maps show that Ophiuchus A consists of three ridges aligned along the north-south direction and that VLA 1623A+B and VLA 1623W are located in adjacent ridges \citep[see Figs. 1 and 4 by][]{Chen2018}.
Therefore, the streamers probed by C$^{18}$O (2--1) which connect VLA 1623A+B with VLA 1623W could be the signature of the gravitational pull across adjacent ridges.
However, the maps by \citet{Chen2018} indicate that the systemic velocity of the two adjacent ridges is similar (both being between +3 and +4 km s$^{-1}$), while our observations indicate that VLA 1623W has a systemic velocity of $+1.6$ km\,s$^{-1}$, which differs from the systemic velocity of VLA 1623AB ($+3.8$ km\,s$^{-1}$) and the two ridges. Thus, this first scenario is unlikely.

\subsection{Hypothesis 2: ejection or flyby}
\label{ejection}

In the second scenario the protostellar source VLA 1623W formed in the same envelope as VLA 1623A and B, and has been later ejected from the multiple system due to a close interaction 
\citep{Murillo2013}.
According to \citet{Harris2018}, the later evolutionary stage of VLA 1623W (Class I) could be due to the loss of most of its original parental envelope during the ejection.
The ejection scenario is further supported by the kinematics of the system.
The ratio between the kinematical and the gravitational energy of the system composed by A, B, and W (protostellar masses of 0.4  M$_{\odot}$, 1.7  M$_{\odot}$, and 0.45 M$_{\odot}$, envelope masses of 0.8  M$_{\odot}$, 0.2  M$_{\odot}$, and 0.04  M$_{\odot}$) is $\simeq 1$, which indicates that the system is likely unstable \citep{Pineda2015}.
Other signatures of instability are: (i) the axis of the circumbinary disk around the A1+A2 binary is misaligned by 12$\degr$ with respect to both the large-scale outflow and the rotation axis of the molecular envelope \citep{Ohashi2022}; (ii) 
the edge-on disk of VLA 1623B counter rotates with respect to the outflow driven by VLA 1623A and the surrounding envelope; and (iii) the circumstellar disks of A1 and A2 have inclinations which may differ by $\sim 70\degr$ based on the orientation of the high-velocity outflows \citep{Harris2018,Murillo2018,Ohashi2022,Codella2022}. 
Given the high dynamic instability, all of the protostars in the system might have been  bound at the time of their formation, with one or more components later ejected due to a close encounter \citep[e.g.,][]{Reipurth2012, Pineda2015}.



In the ejection scenario, the velocity gradient of 1 km\,s$^{-1}$ detected along the southern streamer indicates either material falling on VLA 1623W if the streamer is located in front (i.e. between W and the observer) or  alternatively gas moving away from VLA 1623W towards A+B if the streamer is located on the other side of W.
Note that the close encounter that occurs during the ejection of one of the members of a multiple stellar system has a similar dynamical effect as that of a stellar flyby \citep{Cuello2023}. In the case of a flyby, if the outer perturber, VLA 1623W, follows a prograde orbit near the disk-envelope of A+B this could lead to the formation of streamers like the ones observed in C$^{18}$O \citep[e.g., UX Tau,][]{Menard2020, Zapata2020}. It is, however, puzzling that in VLA 1623 both streamers appear to point towards VLA 1623W since tidal perturbations of the disk typically trigger the formation of two diametrically-opposed streamers, pointing in opposite directions \citep{Clarke1993,pfalzner2003,Cuello2020}.
Interestingly, in the moment 0 maps in Figs. \ref{Con-mom0} and \ref{supermoments} we tentatively detect a spiral arm southern to VLA 1623AB which is diametrically-opposed with respect to the northern spiral arm, N, which connect AB with W. If so, this would indicate that W and AB  interacted recently forming the diametrically opposed spiral arms during the encounter. 
In this scenario the two southern arcs detected in the moment 1 map (S1 and S2) would not be associated to the encounter between W and the AB system.
More in general, assuming that at least one of the streamers was produced by the dynamical interaction during the ejection of VLA 1623W, then the streamers' misalignment with respect to the disks' planes can be due to the fact that misaligned stellar flybys are more likely than coplanar ones \citep{Bate2018, Cuello2019}. 

We stress that both the ejection of a member in young multiple systems as well as stellar flybys are very common processes. In both cases the expected velocities and eccentricities may cover a broad range of values \citep[e.g., ][]{Cuello2023}, therefore, to first order, flybys and ejections leave similar signatures. To distinguish between the two scenarios would require observations at different epochs and accurate astrometry and radial velocity measurements in order to reconstruct the orbit.

\begin{table*}
    \centering
    \begin{tabular}{lllll}
    \hline
     & \multicolumn{2}{c}{Offset between A and W} & \multicolumn{2}{c}{Offset between B and W}  \\
     \hline
   & VLA-X  & ALMA-B6 & VLA-X  & ALMA-B6 \\
   & 1991.8 & 2019.5  & 1991.8 & 2019.5 \\
   \hline
    $\delta$R.A. & $-10\farcs56 \pm 0\farcs03$ &$-10\farcs383 \pm 0\farcs003$ &$-9$\farcs$37 \pm 0\farcs02$ &$-9$\farcs$218 \pm 0\farcs001$ \\
    $\delta$Dec. & $+1\farcs39 \pm 0\farcs03$&$+1\farcs260 \pm 0\farcs003$ &$+1\farcs39 \pm 0\farcs03$ &$+1\farcs120 \pm 0\farcs002$ \\
    $\rho$ &$10\farcs65 \pm 0\farcs06$ &$10\farcs46 \pm 0\farcs02$ & 9$\farcs44 \pm 0\farcs05$& 9$\farcs29 \pm 0\farcs01$ \\
    \hline
    \end{tabular}
    \caption{ The offset (in arcsec) in right ascension ($\delta$R.A.) and declination ($\delta$Dec.), and the  resulting separation ($\rho$) between source A and W and between source B and W at the epochs of the VLA-X observations (1991.8) and the ALMA-Band 6 observations (2019.5).}
     \label{propoermotion}
\end{table*}
\subsection{Proper motions between 1991.8 and 2019.5}
\label{proper motion}
 
In order to {\bf test} the ejection scenario, we estimate the proper motions of VLA 1623W with respect to VLA 1623A and B by combining our FAUST ALMA Band 6 observations of the continuum at 1.3~mm, taken in 2019, with the data at 3.6~cm, taken with the VLA in the X-band in 1991 (\cite{Andre1993}, project AB817).
We measured the separation between A and W, and between B and W, in the VLA-X (1991.8) and ALMA-B6 (2019.5) images, which are separated by 27.7 years. 
Table \ref{propoermotion} reports the offset (in arcsec) in right ascension ($\delta$R.A.) and declination ($\delta$Dec.), and the resulting separation ($\rho$) between source A and source W and between source B and source W  at the two epochs.

From these values, the velocities on the plane of the sky between the sources can be derived. Namely, the velocity on the plane of the sky of A with respect to W is ($-4.4 \pm 1.47$) km s$^{-1}$, while that of B with respect to W is ($-3.5 \pm 1.26$) km s$^{-1}$.
These measurements are in agreement with the proper motions estimated by \cite{Harris2018} based on ALMA observations taken in 2013 \citep{Murillo2013} and 2016 but are affected by a lower uncertainty given the larger distance between the two epochs.
Based on the above estimates, there is no significant difference in tangential velocity between VLA 1623W and A (or B) at the three sigma level. 
However, in the ejection scenario the velocity needed for VLA 1623W to move away from A+B up to a distance of $\sim10\arcsec$ in 10$^4$ years (i.e., the typical Class 0 age) is only 0.65 km\,s$^{-1}$, i.e. below the uncertainty associated with our proper motion measurements ($\sim 1.3-1.5$ km\,s$^{-1}$). 
Thus, the ejection scenario can be neither ruled out nor confirmed by the available proper motion.

\section{Conclusions}
\label{Conclusions}

We report observations of the continuum at 1.3~mm and C$^{18}$O (2--1) line emission towards the multiple protostellar system VLA 1623--2417.
We reveal for the first time the gas associated with the edge-on disk of VLA 1623W. From the gas kinematics, assuming Keplerian rotation, we estimate the source dynamical mass (M$_* = 0.45 \pm 0.08$ M$_{\odot}$).
Moreover, we reveal three streamers connecting VLA 1623W with the VLA 1623A+B system. 
The spatial ($\sim 1300$ au) and velocity ($\sim 2.2$ km\,s$^{-1}$) offset of VLA 1623W with respect to A+B, and its later evolutionary stage (Class I), suggest that either sources W and A+B formed in different cores, or source W has been ejected from the multiple system during its formation, due to the interaction with one of its member.
The available data on proper motions cannot confirm or rule out neither of the two scenarios. Additional kinematical constraints are required in order to test the stellar encounter scenario for VLA 1623.
In addition, observations of shock tracers, such as SO and SiO, at a spatial resolution $\sim 10$ au will allow for verification that the observed streamers are feeding the disk of VLA 1623W, or that vice versa they funnel material from W to A and B. Such observations are required to assess the importance of streamers for disk formation and evolution.


\section*{Acknowledgements}

This project has received funding from the EC H2020 research and innovation
programme for: (i) the project "Astro-Chemical Origins” (ACO, No 811312),  (ii) the European Research Council (ERC) project “The Dawn of Organic
Chemistry” (DOC, No 741002), (iii) the ERC project "Stellar-MADE" (No. 101042275). This study is also supported by grants-in-aid from the Ministry of Education, Culture, Sports, Science, and Technology of Japan (18H05222, 19H05069, 19K14753, and 21K13954), by the Spanish Ministry of Science and Innovation/State Agency of Research MCIN/AEI/10.13039/501100011033 (PID2019-105552RB-C41), “ERDF A way of making Europe”, by the DGAPA PAPIIT grants IN112417 and IN112820, CONACYT-AEM grant 275201, and CONACYT-CF grant 263356, and by
the German Research Foundation (DFG) as part of the Excellence Strategy of the federal and state governments - EXC 2094 - 390783311. 
We are grateful to R.\ Neri for fruitful discussions. D.J.\ is supported by NRC Canada and by an NSERC Discovery Grant. G.B. acknowledges funding from the State Agency for Research (AEI) of the Spanish MCIU trough the PID2020-117710GB-I00 grant funded by MCIN/AEI/10.13039/501100011033. We thank the anonymous referee for the very constructive comments and suggestions.

\section*{Data Availability}
The raw data will be available on the ALMA archive at the end of the proprietary period (ADS/JAO.ALMA\#2018.1.01205.L).



\bibliographystyle{mnras}
\bibliography{example} 
\noindent




%
\appendix

\section{Channel maps of C$^{18}$O ($2-1$) towards source W at low velocities}

\begin{figure*}
\centering
\vspace{-2cm}
\includegraphics[width=12.5cm, angle =90]{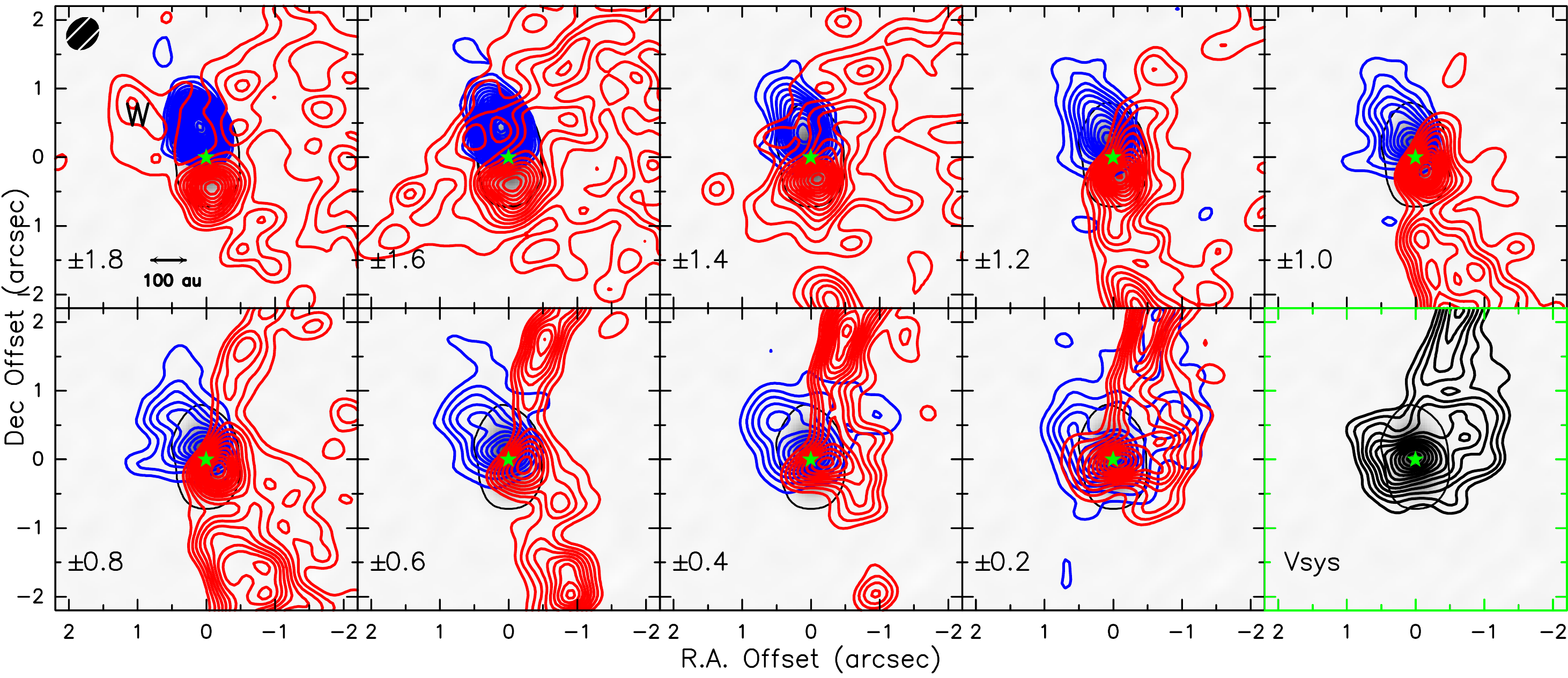}
\vspace{-2cm}
    \caption{Channel maps of C$^{18}$O (2--1) emission towards the VLA 1623W protostar (green star) at low red- and blue-shifted velocities, i.e. up to $\pm1.8$ km s$^{-1}$ with respect to the VLA 1623W systemic velocity ($V_{\rm sys}$(W) $\sim +1.6$ km s$^{-1}$).  First contours and steps are 3$\sigma$ (7.7 mJy beam$^{-1}$). The velocity offset with respect to $V_{\rm sys}$(W) is reported in the bottom left corner of each panel. The black contour is the 3$\sigma$ level of the 1.3~mm continuum emission, which is also shown by the gray scale background. The synthesized beam (0$\farcs$48 $\times$ 0$\farcs$40) is shown in the top left corner of the first channel. }
    \label{appendix}
\end{figure*}

Figure \ref{appendix} shows the channel maps of C$^{18}$O (2--1) emission towards source W at low-velocities, i.e. up to  $\pm1.8$ km s$^{-1}$  with respect to the systemic velocity of W ($V_{\rm sys}$(W) $\sim +1.6$ km s$^{-1}$). 
The maps show that the kinematics of the gas in the disk, which is well detected at high-velocities (see Fig. \ref{channel_small}), is affected by the extended emission from the streamers and/or the residual envelope at low velocities.

\bsp	
\label{lastpage}
\end{document}